\documentclass[reprint,aps,prb,amsmath,amssymb,onecolumn,showkeys, superscriptaddress,floatfix]{revtex4-2}
\usepackage{graphicx}
\usepackage{dcolumn}
\usepackage{bm}
\usepackage[colorlinks = true, allcolors = blue]{hyperref}
\usepackage{verbatim}
\usepackage{soul}
\usepackage{float}
\usepackage{xcolor}
\usepackage{multirow}
\usepackage{tabularx}
\usepackage{orcidlink}

\begin{document}
\title{Tuning of quantum nanoscaled friction within the Prandtl-Tomlinson model}

\author{Dai-Nam Le\,\orcidlink{0000-0003-0756-8742}}
\email{dainamle@usf.edu}
\homepage{https://sites.google.com/view/dai-nam-le/}
\affiliation{Department of Physics, University of South Florida, Tampa, Florida 33620, USA}

\author{Lilia M. Woods\,\orcidlink{0000-0002-9872-1847}} 
\email{lmwoods@usf.edu}
\homepage{https://www.amd-woods-group.com/}
\thanks{Corresponding authors}
\affiliation{Department of Physics, University of South Florida, Tampa, Florida 33620, USA}

\date{\today}

\begin{abstract}
Nanoscaled  is a fundamental tribological phenomenon with complex behavior of its dynamical force. Here, we utilize the Prandtl-Tomlinson framework to  investigate systematically the different means of control of the frictional force at the quantum and classical levels. It is found that the frictional dynamics can be controlled by the corrugation and characteristic length ratio parameters dependent upon properties of the nanoparticle-chain system. In addition to the stick-slip regime, other types of motion are uncovered, highlighting the richness of the frictional dynamics. The importance of Landau-Zener tunneling for the quantum motion is also analyzed. These findings provide valuable insights for interpreting experimental observations and controlling quantum frictional behavior in nanoscale systems.

\begin{figure}[H]
    \centering
    \includegraphics[width=0.6\textwidth]{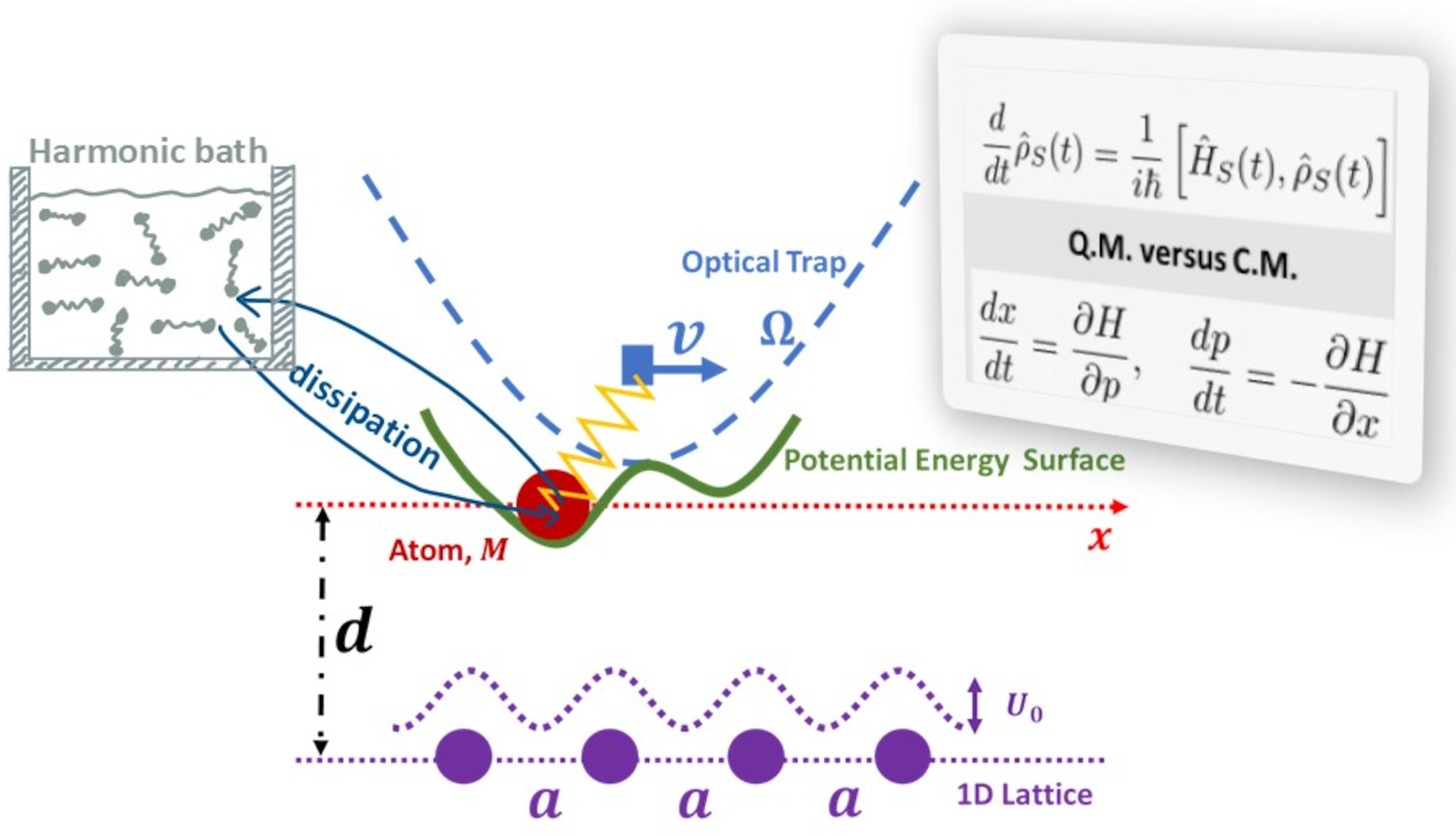}
\end{figure}
\end{abstract}

\maketitle

\section{\label{sec:intro}Introduction}

Relative motion between objects generates friction associated with energy dissipation in the environment. Friction has been studied extensively, although a complete understanding of the underlying mechanisms and their material-dependent aspects is not fully available \cite{Vakis2018, Volokitin2007}. While non-contact mechanisms such as Casimir friction can also arise from electromagnetic fluctuations between bodies in relative motion \cite{Volokitin2007, Milton2016, Decca2020, Lombardo2021, Modi2026}, contact friction remains the dominant and most technologically relevant form of energy dissipation. Contact friction, especially in the context of materials, is important for optimizing the operation of various devices, including applications using force microscopy setups \cite{Szlufarska2008, Wang2022}. 

First principles computational studies \cite{Wolloch2019,Cahangirov2012,Restuccia2023, Losi2023, Woods2024, Silvestrelli2025,Dang2025} have focused on adhesion and corrugation properties in a variety of materials seeking to identify systems for targeted tribological applications\cite{Wolloch2018,Torche2022, Kajita2023}. On the other hand, the Prandtl–Tomlinson model  is typically used for the investigation of dynamic contact friction \cite{Tomlinson1929,Prandtl1928}. This classical framework describes a particle, lacking internal degrees of freedom, that moves above an atomic chain modeled as a periodic potential. The frictional process is typically described based on empirical material-dependent parameters for corrugation, velocity, and particle-chain interaction. The relative motion of atomic and/or molecular chains has also been studied within the classical Frenkel–Kontorova model, which has been applied to study chain dislocations, nonlinearities, and topological defects \cite{Krajewski2004,Krajewski2005,Xu2007}. Monte Carlo simulations and path integral calculations \cite{Bustos2016,Furlong2009,Nabulsi2020} have indicated that classical models are not sufficient, and quantum mechanical effects must be included to better understand frictional processes. 

Recently, the Prandtl-Tomlinson
model has been broadened to take into account quantum mechanical effects in the energy dissipation and frictional force in a moving nanoparticle above an atomic chain \cite{le2025quantum, Zanca2018}. An essential aspect in this theory is the Landau-Zener (LZ) tunneling due to avoided energy level crossing, a contributing factor to the stick-slip nanoparticle motion \cite{Landau1932,Zener1932}. The LZ theory has been applied to various dissipative phenomena modeled via coupling to an external bath harmonic oscillators with a constant environmental temperature \cite{Pokrovsky2002,Huang2018,Arceci2017}. In the stick-slip frictional regime the LZ tunneling is responsible for reduced dissipation and a smaller retarding force in the quantum Prandtl-Tomlinson model compared to the classical situation \cite{le2025quantum, Zanca2018}. Contact friction at the nanoscale is typically modeled as a dissipative phenomenon by utilizing master equation dynamic approaches taking into account the interaction with the external environment. Consequently, nanoscaled friction is a part of a wide range of problems for which not only dissipation, but decoherence and fluctuational dynamics also become important in transport and nonequilibrium processes \cite{Bandyopadhyay2020,Bandyopadhyay2021,Pei-Yun2017}.

Recently, it was realized that the stick-slip motion is directly related to the corrugation parameter $\eta$, a unitless constant capturing the ratio of various physical characteristics of the nanoparticle/atomic chain system \cite{le2025quantum}. In particular, the stick-slip regime was studied for  $\eta\sim 1$, where LZ tunneling may be prominent. However, since $\eta$ is a collection of several parameters in the model, motion beyond the stick-slip regime is also captured in the Prandtl-Tomlinson model. In fact, by varying the different physical properties we uncover a complex frictional behavior in nanoscale systems controlled not only by the corrugation parameter at the quantum level, but also by the relative length between the atomic chain period and the characteristic length of the optical trap. Depending on the physical properties of particle/chain system, it is possible to have motion that is stick-slip-free or multiple stick-slip as opposed to the single stick-slip motion as found for $\eta \sim 1$ case \cite{le2025quantum, Zanca2018}.

In this study, the frictional properties of a nanoparticle moving above an atomic chain are examined in terms of various parameters in the Prandtl–Tomlinson model. We identify criteria which control the emergence of different regimes of motion, in addition to stick-slip. This type of frictional tuning is presented at the classical and quantum mechanical level. In the quantum formalism, kinetic and dissipative properties are studied by employing  density matrix formalism in the Liouville-von Neumann equation in   the Markov approximation. In the classical formalism, the time evolution of these properties is determined by solving the stochastic Newton equation. The dependence of the lateral force upon properties in the model provides a qualitative understanding of nanoscaled contact friction and it highlights differences between quantum and classical levels of description. An important aspect of our study is the realization that the presence of LZ tunneling is not always associated with stick-slip frictional motion. The realization of dynamical phases with no stick-slip or multiple stick-slip events and their tunability with the physical parameters of the particle-chain system significantly broadens our understanding of contact nanoscaled friction beyond currently available studies \cite{le2025quantum, Zanca2018}.

\section{\label{sec:2}Theoretical Model of the nanoscaled friction}

\begin{figure}[!ht]
    \centering
    \includegraphics[width=0.55\textwidth]{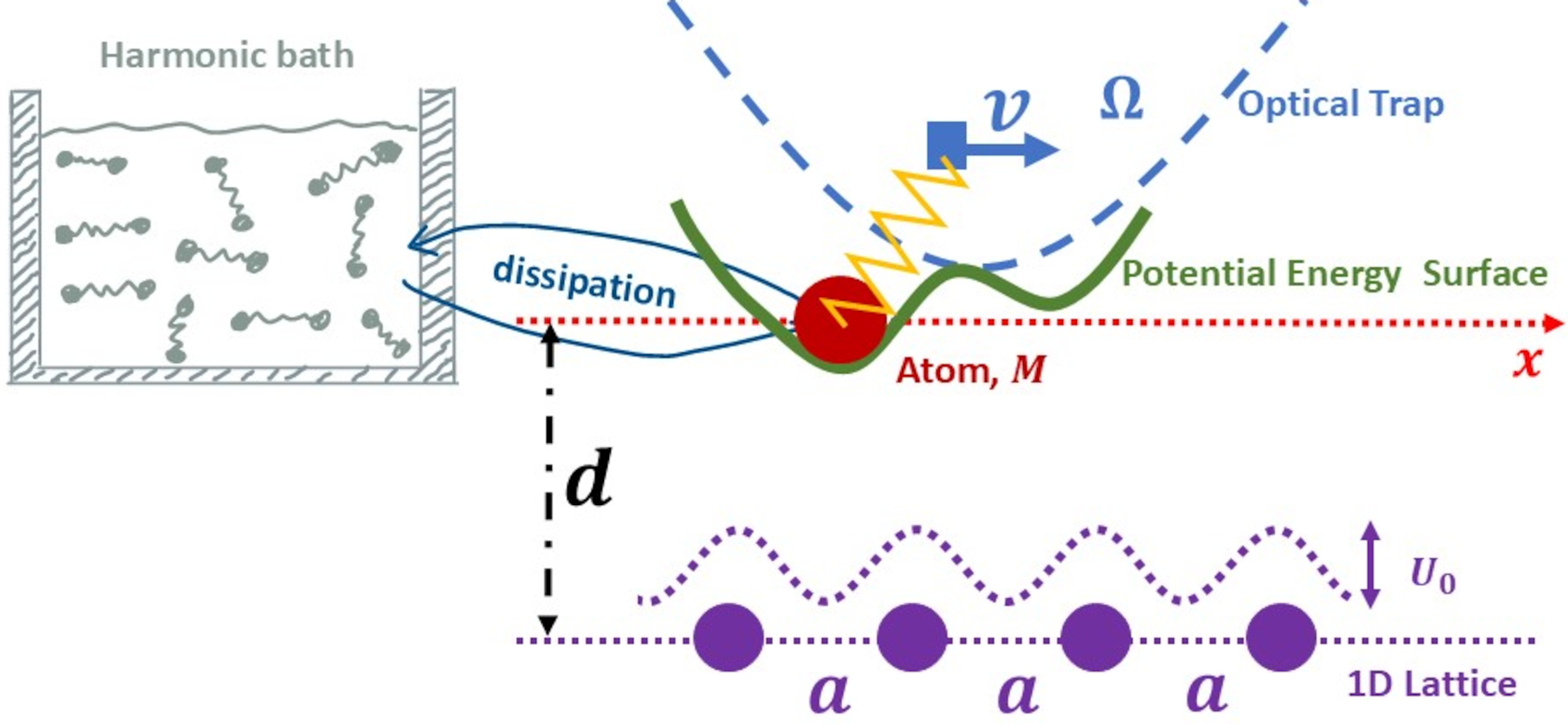}
    \caption{\label{fig:1}Graphical sketch of the considered system where a nanoparticle confined by a parabolic optical trap moves above an one-dimensional chain of atoms. The coupling to an external thermal bath results in frictional dissipation.
    }
\end{figure}

The system under consideration consists of a particle with mass $M$ whose motion above a one-dimensional chain with a lattice constant $a$ is facilitated by a moving harmonic trap with characteristic frequency $\Omega$ and velocity $v$, as shown in Figure \ref{fig:1}. The frictional aspect of this motion is modeled as dissipation due to coupling of the particle to a reservoir of harmonic oscillators with an environmental temperature $\mathcal{T}$ as described in the Caldeira–Leggett model \cite{Caldeira1981, Weiss2012, Zanca2018,le2025quantum}. The full quantum mechanical Hamiltonian consists of two pieces describing the particle/chain system $\hat{H}_S (t) $ and the interaction with the environment $\hat{H}_{int+B} (t) $, such that 
\begin{eqnarray}
    \label{eqn:H-full}
    \hat{H}(t) &=& \hat{H}_S (t) + \hat{H}_{int+B}, \\
    \label{eqn:S}
    \quad  \hat{H}_S (t) &=& \frac{1}{2M} \hat{p}^2 + \frac{1}{2} M \Omega^2 \left( x - v t \right)^2 + U_0(d, a) \sin^2 \left( \frac{\pi}{a} x \right) ,\\
    \label{eqn:HQ}
    \hat{H}_{int+B} &=&  \sum_i \left[ -\frac{1}{2 m_i} \frac{\partial^2}{ \partial x_i ^2}  + \frac{1}{2} m_i \omega_i^2 \left( x_i - \frac{c_i}{m_i \omega_i^2} \hat{A} \right) ^2 \right] .
\end{eqnarray}
The first term in Equation \eqref{eqn:S} corresponds to the kinetic energy of the moving particle taken into account via the momentum operator $\hat{p}$, and the second term is the time-dependent potential energy of the trap. Also, $U_0 (d,a)$ is the interaction potential between the particle and the chain, which includes short-ranged and long-ranged coupling. As shown in Reference  \cite{le2025quantum}, the potential correlates with  the lattice constant $a$ via $U_0(d,a) \propto a^{-1}$, while its correlation with the separation $d$ has a more complicated form $U_0(d,a)\sim C_{sr} \exp{(-d/d_0)} - C_{lr}/d^{-\sigma_{lr}}$ where $C_{sr}$ and $C_{lr}$ are short- and long-range interatomic interaction constants while $\sigma_{lr}$ is the power law of the long-range interatomic interaction (typically $\sigma_{lr} \approx 6$ according to the Born-Mayer-Buckingham model \cite{Stone2013} but it could be varied due to many-body or nonlinear effects \cite{Le2024, Dang2025CPC}).

The frictional dissipation is modeled according to the Caldeira–Leggett model \cite{Caldeira1981, Weiss2012, Zanca2018,le2025quantum} with the term $\hat{H}_{int+B}$ describing the coupling of the particle/chain system to an external thermostat of harmonic oscillators. In this Hamiltonian, $x_i$, $m_i, \omega_i, c_i$ are the spatial degrees of freedom of the bath oscillator modes with their position, mass, frequency, and interaction strength, respectively. The renormalization operator $\hat{A}$ is chosen as $\hat{A} = \sin \left( \frac{2\pi x}{a} \right)$ reflecting the global symmetry of the Hamiltonian $\hat{H}_S (t)$. 

The dissipation is defined by the spectral function $J(\omega) = \sum\limits_i \frac{c_i^2}{2 m_i \omega_i} \delta \left( \omega - \omega_i \right)$ \cite{Dittrich1998, Weiss2012}. In the Ohmic dissipation limit, considered here,  $J(\omega) = 2 \alpha \omega e^{-|\omega| / \omega_c}$, where $\alpha$ is a dimensionless constant representing the coupling between the particle/chain system and the harmonic bath. The cutoff frequency $\omega_c$ separates the low-frequency regime for which $J(\omega) \propto \omega$ and the high-frequency regime specified by the Ohmic damping. One notes that when $\alpha = 0$, $J(\omega) = 0$ for all frequencies and $c_i = 0$, i.e., there is no coupling between the system and the bath and $\hat{H}_{int+B} = 0$. The periodicity of $T=a/v$ of the system is also preserved in the full Hamiltonian in \eqref{eqn:H-full}. 


The dynamic evolution of the system obeys the quantum Liouville-von Neumann master equation for the density matrix operator \cite{Yamaguchi2017}. It is assumed that the Born-Markov approximation is valid, which implies   $\alpha\ll1$. Under this approximation, the environment is  unaffected by the system, and the external bath has very short correlation times compared to the relaxation time of the system meaning that memory effects are not important.

Thus, we have \cite{le2025quantum}
\begin{eqnarray}
    \label{eqn:quantum-1}
    \dfrac{d}{dt} \hat{\rho}_S (t) = - \frac{i}{\hbar} \left[ \hat{H}_S(t) + 2 \alpha \hbar \omega_c \hat{A}^2, \hat{\rho}_S (t) \right] - \left\{ \left[ \hat{A}, \hat{S} (t) \hat{\rho}_S(t) \right] + \text{Hermitian Conjugate} \right\}.
\end{eqnarray}
The bath-convoluted operator $\hat{S}(t) \approx \sum\limits_{m,n} \left< p(t) \right| \hat{A} \left| p' (t) \right> \Gamma (E_{p'} - E_p) \left| p (t) \right> \left< p'(t) \right| $ is related to the eigenenergies $E_p (t)$ and eigenstate $\left| p(t) \right>$  of the renormalized Hamiltonian $\hat{H}_S (t) + 2 \alpha \hbar \omega_c \hat{A}^2$. The bath-induced transition rate $\Gamma (E) = \int_0^{+\infty} C(\tau) e^{i \hbar^{-1} (E + 0^+) \tau} d \tau$ is the Fourier transform of the bath correlation function $C (\tau) = \int_{-\infty}^{+\infty} e^{i \omega \tau} f_{BE} (\omega) J(\omega) d \omega $ where $f_{BE} (\omega) = \left( e^{\hbar \omega / k_B \mathcal{T}} - 1 \right)^{-1}$ is the Bose-Einstein statistics. The solution of Equation \eqref{eqn:quantum-1} can be sought in a matrix form using the eigenstates $\left| p(t) \right>$. For our subsequent calculations we use $\alpha=2\times10^{-3}$,  $\omega_c=50$ $\Omega$, and $T=40\pi/\Omega$ which guarantee the positivity of $\hat{\rho}$ and the Markovianity of the system. Also, the Hamiltonian of the system is taken in terms of a  time-dependent simple harmonic oscillator basis set $\left| p^{(0)} (t) \right> = \varphi^{SHO}_p \left( x - v t \right)$ with exact matrix elements 
 \cite{le2025quantum},
 \begin{eqnarray}
\label{eqn:quantum-2}
   && ( H_S )_{p,p'} (t) = \left( p + \frac{1}{2} + \eta \bar{\Lambda}^2 \right) \delta_{p,p'} + \eta \bar{\Lambda}^2 \mathcal{V}_{p,p'}(t), \\
\label{eqn:quantum-3}
   && \mathcal{V}_{p,p'}(t) = (-1)^{\left\lfloor \frac{p - p'}{2} \right\rfloor} \sqrt{ \frac{\text{min}(p,p')!}{\text{max}(p,p')!}} e^{-\frac{\bar{\Lambda}^{-2}}{4}} \left( \frac{\bar{\Lambda}^{-2}}{2 } \right)^{\frac{|p-p'|}{2}} \mathcal{L}_{\text{min}(p,p')}^{|p-p'|} \left( \frac{\bar{\Lambda}^{-2}}{2} \right) \times \left\{ \begin{matrix}
        + \sin \left( \frac{2\pi t}{T} \right) & |p-p'| \text{ is odd} \\
        - \cos \left( \frac{2\pi t}{T} \right) & |p-p'| \text{ is even}
    \end{matrix} \right..
\end{eqnarray}
where $\mathcal{L}_{\text{min}(p,p')}^{|p-p'|}$ are the associated Laguerre polynomials \cite{gradshteyn2014table}. All of these quantities and subsequent solutions are explicitly dependent on the {\bf corrugation parameter} $\eta = \frac{2\pi^2 U_0}{M \Omega^2 a^2}$ and the {\bf length ratio} $\bar{\Lambda} = \frac{a}{2\pi \ell}$, where $\ell = \sqrt{\hbar/ M\Omega}$ is the characteristic length of the harmonic trap. The inherent periodicity in the system is also present via  {\bf period} $T=\frac{a}{v}$. It is these three parameters, $\eta, \bar{\Lambda}$, and $T$ that control the qualitative behavior of the dynamics. The numerical diagonalization of $ \hat{H}_S$ is obtained using the fourth-order Runge-Kutta method, which enables the numerical expressions for the eigenvalues $E_p(t)$ and eigenfunctions $\left| p(t) \right>$ (see Reference  \cite{le2025quantum} for more details). The initial density matrix for the ground state is set as $\hat{\rho}_S(0) = \left| 0 (0) \right> \left< 0(0) \right|$. We note that although the Hamiltonian and density matrices are infinite, in the numerical calculations their size is finite. Details about the basis cutoff $N_{\text{max}}$ are given in Appendix \ref{appC}.

The main interest in our study is the frictional force on the nanoparticle and its dependence upon the parameters in the problem. The frictional force is lateral and it opposes the direction of the motion. In the quantum mechanical limit, it is related to the trace of the density matrix. We are interested in its maximum value, which is defined by the following relation, 
\begin{eqnarray}
    \label{eqn:quantum-lateral}
    \frac{\left<F_L \right>_{\text{max}}^{qm}}{F_{0}} = \dfrac{2\pi}{\eta} \text{Tr} \left[ \left(\frac{t}{T} - \frac{\hat{x}}{a}\right) \hat{\rho}_S (t) \right]_{\text{max}} ,
\end{eqnarray}
where $F_{0}={\pi U_0}/{a}$. The lateral force is a time-dependent quantity and here we are tracking its maximum value in the $t/T=[0,3]$ interval. The maximal lateral force is experimentally accessible, either by atomic force or scanning tunneling microscopy (AFM or STM) \cite{ Socoliuc2004, Bartels1997, Liu2015, Almeida2016, Yang2018} or optical means \cite{Erik2007,Gangloff2015,Counts2017}.


Auxiliary quantities that help with the analysis of the results for the force are the minimum population of the instantaneous ground state and the maximum linear entropy the system can take. These also depend on the density matrix and can be evaluated as 
\begin{equation}
    \label{eqn:quantum-minP}
    P^\text{min }_{p = 0} = \left< 0 (t) \middle| \hat{\rho}_S (t) \middle| 0 (t) \right>_\text{min }, \quad
    S^\text{max }_L = \frac{N_{\text{max}}}{N_{\text{max}}-1} \left[1 -  \text{Tr} \left( \rho_S^2(t) \right)_\text{min } \right]. 
\end{equation}
The quantity $ P^\text{min }_{p = 0}$ is related to the occurrence of LZ diabatic transitions involving the ground state as shown in earlier works \cite{Zanca2018, le2025quantum}. It was found that LZ transitions between the ground state and the first excited state become possible for $ P^\text{min }_{p = 0} < 0.5$, otherwise, the populations simply oscillates between these two states without any tunneling. The maximal linear entropy $S^\text{max }_L$, on the other hand, is connected to the role of temperature. Larger  $S^\text{max }_L$ indicates larger disorder in the system due to the stronger influence of the thermal bath \cite{le2025quantum}. The appearance of the renormalization factor ${N_{\text{max}}}/{(N_{\text{max}}-1)}$ is due to the finite cutoff size $N_{\text{max}}$ of the Hamiltonian $\hat{H}_S$, as shown in Ref.  \cite{Peters2004}. Notably, for shorter times the quantum motion is determined primarily by the quantum states of the system, their superposition and tunneling events due to LZ diabatic transitions. In the long-time limit, however, the system experiences a complete loss of coherence due to the Ohmic bath from the environment.

The nanoscaled friction is also solved within the classical Prandtl-Tomlinson model. For this purpose, the classical Hamiltonian (see Equations \eqref{eqn:H-full}-\eqref{eqn:HQ}) and the associated Hamilton equations of motion yield the stochastic Newton's equation  \cite{Weiss2012, Wilkie2005, le2025quantum} 
\begin{equation}
    \label{eqn:classical}
    \dfrac{d^2 (x/a) }{d (t/T)^2} = - \Omega ^2 T^2 \left[ \frac{x}{a} - \frac{t}{T} + \dfrac{\eta}{2\pi} \sin \left( \dfrac{2\pi x}{a} \right) \right] - \frac{2 \pi \alpha \Omega T}{ \bar{\Lambda}^{2}} \cos^2 \left( \dfrac{2\pi x}{a} \right) \dfrac{d (x/a) }{d (t/T)} + \xi_{ran} \left(t/T\right),
\end{equation}
where the random force $\xi_{ran} \left(t/T\right)$ obeys the fluctuation-dissipation theorem with a two-time correlator $\Big< \xi_{ran} \left(\dfrac{t}{T}\right) \Big. \allowbreak \Big.  \xi_{ran} \left(\dfrac{t'}{T}\right) \Big> = \dfrac{\alpha (\Omega T)^3}{2\pi \bar{\Lambda}^4} \dfrac{k_B \mathcal{T}}{\hbar \Omega} \cos^2 \left( \dfrac{2\pi x}{a} \right) \delta \left( \dfrac{t}{T} - \dfrac{t'}{T} \right)$ \cite{Rytovbook, Weiss2012}. The initial condition of the nanoparticle is set as $x(0) = 0, v(0) = 0$.

From the solution of the above equation, the maximum value of the lateral force between the harmonic trap and the particle is given as
\begin{eqnarray}
    \label{eqn:classial-lateral}
    \frac{\left<F_L \right>_{\text{max}}^{cl}}{F_{0}} = \frac{1}{N_{\text{ran}}} \sum\limits_{j=1}^{{N_{\text{ran}}}} \dfrac{2\pi}{\eta} \left( \frac{t}{T} - \frac{x_j}{a} \right)_{\text{max}},
\end{eqnarray}
where $x_j(t)$ is solution of Equation \eqref{eqn:classical} at $j^{th}$ random run and $N_{\text{ran}}$ is the total random runs of Equation \eqref{eqn:classical}. Details about $N_{\text{ran}}$ are given in Appendix \ref{appC}.

The solutions to Equations \eqref{eqn:quantum-1}-\eqref{eqn:quantum-3} for the quantum case and Equation \eqref{eqn:classical} for the classical case are controlled by the $\eta, \bar{\Lambda},$ and $T$ parameters. As a guiding framework, we present several analytical expressions for the lateral force in the quantum and classical cases, neglecting coupling to the dissipative heat bath, which is valid as long as the dimensionless coupling strength $\alpha$ and the bath temperature $k_B \mathcal{T}$ are small enough i.e. $\alpha \ll 1$ and $k_B \mathcal{T} \ll \hbar \Omega$. In particular, the quantum lateral force takes the asymptotic form 
\begin{eqnarray}
\label{eqn:quantum-lateral-approx}
\frac{\left<F_L \right>_{\text{max}}^{qm}}{F_{0}} & \approx & \left\{ \begin{array}{ll}
       \frac{2\pi}{\eta \Omega T} + \exp\left( - \frac{\bar{\Lambda}^{-2}}{4} \right)  & \text{for } \eta < 1 \text{ or } \eta > 1 \text{ and } \bar{\Lambda} \ll \left(\eta + \sqrt{\eta^2 - u_{\eta}^2}-u_{\eta}^2/2\right)^{-1/2}, \\
      \text{const} < 1  & \text{for } \eta \gtrsim 1 \text{ and } \bar{\Lambda} \gtrsim \left(\eta + \sqrt{\eta^2 - u_{\eta}^2}-u_{\eta}^2/2\right)^{-1/2}
    \end{array} \right. ,
\end{eqnarray}
while the classical lateral force can be expressed as,
\begin{eqnarray}
\label{eqn:classical-lateral-approx}  \frac{\left<F_L \right>_{\text{max}}^{cl}}{F_{0}} & \approx & \left\{ \begin{array}{ll}
       \frac{2\pi}{\eta \Omega T} + 1 & \text{for } \eta < 1, \\
       1  & \text{for } \eta \gtrsim 1,
    \end{array} \right..
\end{eqnarray}
Here $u_{\eta}$ is the solution of the equation $u = \eta \sin u$ in the $(0,\pi)$ interval (see Appendices \ref{appA} and \ref{appB} for details). From the above expressions, we notice that the classical lateral force depends solely on the corrugation parameter $\eta$, which is consistent with the fact that there is no $\hbar$ in $\eta$. However, both $\eta$ and  $\bar{\Lambda}$ affect the quantum lateral force as the former one determines the shape of potential energy surface while the latter one captures the quantum nature of the system through the appearance of $\hbar$ in $\bar{\Lambda}$.

\section{\label{sec:3}Regimes of frictional motion}

The above described framework gives the tools to study the frictional force within the Prandtl-Tomlinson model at the quantum mechanical and classical levels. Numerical calculations are obtained by solving Equations \eqref{eqn:quantum-1} and \eqref{eqn:classical} and the analytical expressions in Equations \eqref{eqn:quantum-lateral-approx}-\eqref{eqn:classical-lateral-approx} can be used to validate the numerical solution and for deeper analysis of the results. These analytical expressions also show how the lateral force depends on the physical characteristics in the Prandtl-Tomlinson model. In particular, we find that the frictional motion can be characterized with different feature based on $\eta < 1$, $  \eta \sim 1$ and $\eta>1$. Recent studies \cite{le2025quantum,Zanca2018} have focused on the stick-slip motion with $\eta\sim 1$. In this case, LZ diabatic transitions result in a reduced quantum friction when compared to the classical motion where such tunneling effects are not feasible \cite{Gnecco2012,Muser2011}. 

However, we find richer parameter-dependent motion based on the composite parameters $\eta$ and $\bar{\Lambda}$. These can be tuned by modulating $a, U_0, \Omega$, etc. in ways that are not necessarily independent. For example, the relation $u_0={U_0}/{\hbar\Omega} = 2 \eta \bar{\Lambda}^2$ shows a correlation between $\bar{\Lambda}$ and $U_0, \Omega$. Manipulating $\Omega$ and $a$ such that $\Omega \sim a^{-3/2}$ keeps $\eta$ constant while changing $\bar{\Lambda} \sim a^{1/4}$ (see definitions $\eta=\frac{2\pi^2 U_0}{M\Omega^2a^2}$ and $\bar\Lambda=\frac{a}{2\pi}\sqrt{\frac{M\Omega}{\hbar}}$). Also, adjusting the separation distance $d$ changes the periodic potential $U_0 = U_0 (d,a)$ \cite{le2025quantum}, which affects $\eta$, but not $\bar{\Lambda}$.

\subsection{\label{sec:Lambda}Modulations by changing the relative length ratio $\bar{\Lambda}$}

\begin{figure}[!ht]
    \centering
    \includegraphics[width=0.8\linewidth]{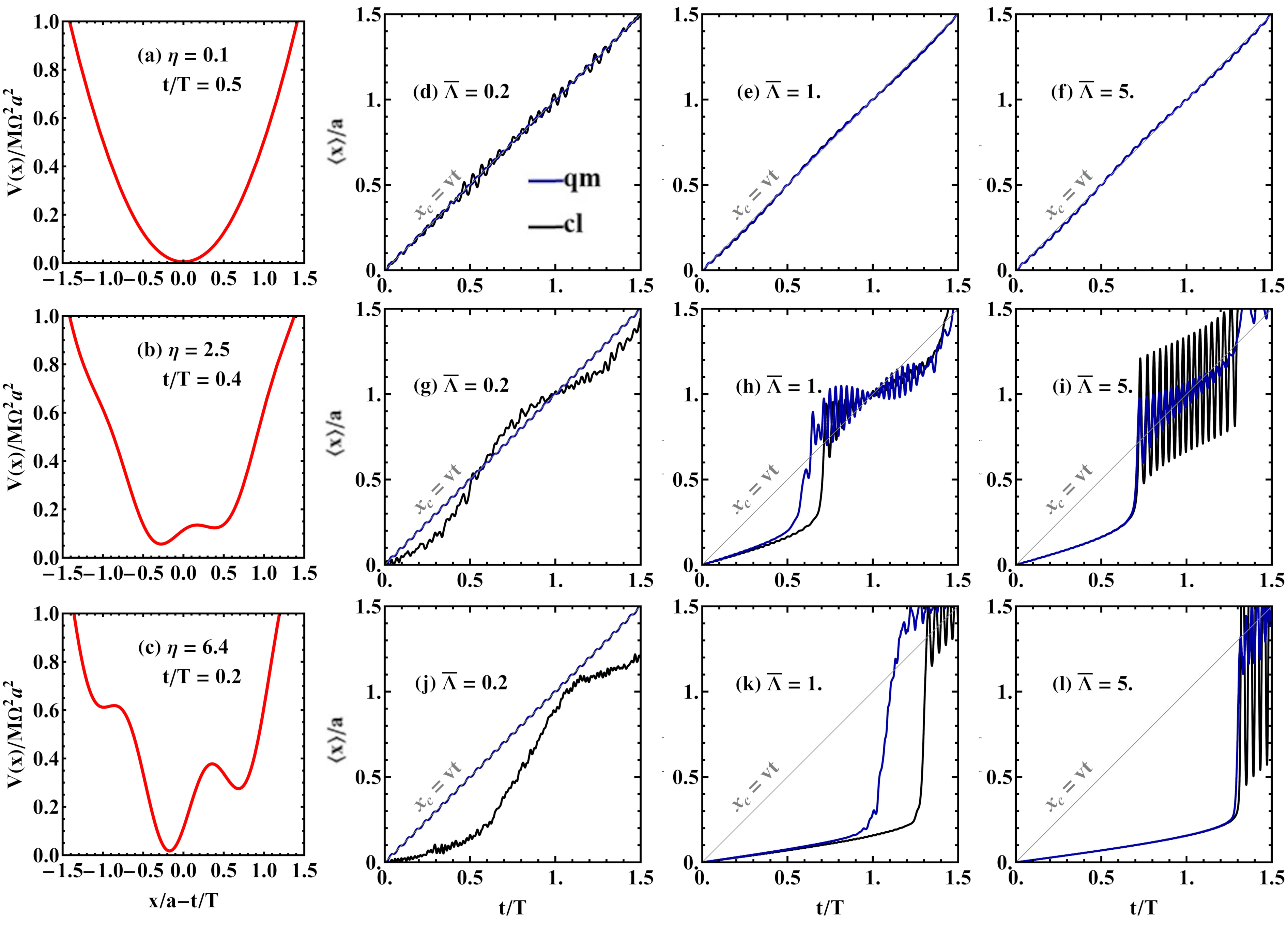}
    \caption{\label{fig:2}Total potential energy surface $V(x)$ normalized by $M\Omega^2a^2$ as a function of $x/a-t/T$ for: (a) $\eta=0.1$ at $t/T=0.5$; (b) $\eta=2.5$ at $t/T=0.4$; (c) $\eta=6.4$ at $t/T=0.2$. Quantum ${\left<x\right>}/{a} = \text{Tr }\left( {\hat{x}}\hat{\rho}_S(t) \right)/a$ and classical  ${\left<x\right>}/{a} = \left({1}/{N_{\text{ran}}}\right) \sum\limits_{j=1}^{N_{\text{ran}}} x_j / a $ trajectories as a function of $t/T$ for: $\eta=0.1$ at (d) $\bar{\Lambda} = 0.2$, (e) $\bar{\Lambda} = 1$, (f) $\bar{\Lambda} = 5$; $\eta=2.5$ at (g) $\bar{\Lambda} = 0.2$, (h) $\bar{\Lambda} = 1$, (i) $\bar{\Lambda} = 5$; and $\eta=6.4$ at (j) $\bar{\Lambda} = 0.2$, (k) $\bar{\Lambda} = 1$, (l) $\bar{\Lambda} = 5$. The displacement of the trap  $x_c=vt$ is also shown.  The bath parameters are $\alpha = 2 \times 10^{-3}$, $\omega_c = 50$ $\Omega$ and $k_B \mathcal{T} = 0.1$ $\hbar\Omega$.}
\end{figure}

\begin{figure}[!ht]
    \centering
    \includegraphics[width=0.90\textwidth]{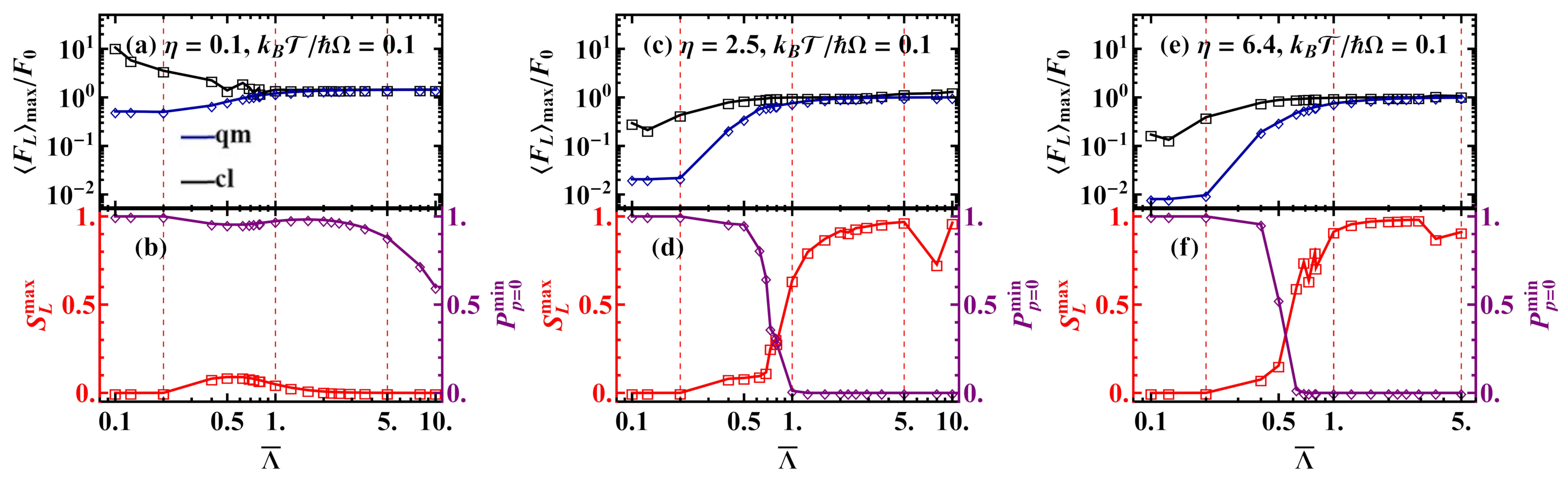}
    \caption{\label{fig:3} Classical (black) and quantum (blue) maximal lateral forces scaled by $F_0={\pi U_0}/{a}$ as a function of $\bar{\Lambda}$ for: (a) $\eta=0.1$; (c) $\eta=2.5$; (e) $\eta=6.4$. Maximal linear entropy $S_L^{\text{max}}$ (red) and minimal ground state  population $P_{p=0}^{\text{min}}$ (purple) as a function of $\bar{\Lambda}$ for: (b) $\eta=0.1$; (d) $\eta=2.5$; (f) $\eta=6.4$.  The bath parameters are $\alpha = 2 \times 10^{-3}$, $\omega_c = 50$ $\Omega$ and $k_B \mathcal{T} = 0.1 $ $\hbar\Omega$. Also, $N_{\text{ran}}=200$ for all of the classical calculations; $N_{\text{max}} = 200$ for all quantum calculations in (a), (b) with $\bar{\Lambda} \leq 5$ and (c) with $\bar{\Lambda} \leq 3.2$, $N_{\text{max}} = 500$ for quantum calculations in (b) with $\bar{\Lambda} = 8$ and $10$ and $N_{\text{max}} = 650$ for quantum calculations in (c) with $\bar{\Lambda} = 5$.}
\end{figure}

As a first step towards understanding the parameter dependence of the frictional motion, in Figure \ref{fig:2} we show the potential energy surface and the average displacement $\frac{\left<x\right>}{a} = \text{Tr }\left( \frac{\hat{x}}{a}\hat{\rho}_S(t) \right)$ (quantum) and $\frac{\left<x\right>}{a} = \frac{1}{N_{\text{ran}}} \sum\limits_{j=1}^{N_{\text{ran}}} \frac{x_j}{a}$ (classical) of the system in the three regimes for the representative examples of $\eta=0.1, 2.5$ and $6.4$.  In Figure \ref{fig:3}, results are shown for the lateral force as a function of $\bar{\Lambda}$ for the same values of $\eta$.

For $\eta < 1$, the particle is driven along a potential surface with only one global minimum, as shown in Figure \ref{fig:2}(a) for $\eta=0.1$. There are no LZ diabatic transitions, and in both, the classical and quantum motions, shown in Figures \ref{fig:2}(d-f), the particle just oscillates between the  ground state ($p = 0$) and the first excited state ($p = 1$) around the center of the driven optical trap $x_c = vt$. Figure \ref{fig:3} (b) confirms that for smaller $\bar{\Lambda}$ the level population is $P_{p = 0}^{\text{min}}\sim 1$.  For larger $\bar{\Lambda}$, $P_{p = 0}^{\text{min}}$ decreases due to the admixture with the $p=1$ level, but it is always larger than a half, $P_{p = 0}^{\text{min}}> 1/2$.  The entropy is zero everywhere with a $S_L<0.05$ region around $\bar{\Lambda}\sim 0.5$.

Figure \ref{fig:3} (a) shows that the  classical force is found to be almost a constant with respect to $\bar{\Lambda}$ \textcolor{red}{i.e. $\left< F_L \right>_{\text{max}}^{cl} \approx 1.44 F_0$}, consistent with the analytical expression in Equation \eqref{eqn:classical-lateral-approx} where \textcolor{red}{$\left< F_L \right>_{\text{max}}^{cl} = 1.5 F_0$}. The behavior of the quantum lateral force, also shown in Figure \ref{fig:3} (a)  agrees with the analytical expression in Equation \eqref{eqn:quantum-lateral-approx} ($\left< F_L \right>_{\text{max}}^{qm} \approx 0.5 F_0$ at $\bar{\Lambda} = 0.2$ for example), and it can be understood by examining its difference with the classical lateral force, found as 
\begin{equation}
    \label{eqn:difference-1}
    \frac{\left<F_{L}\right>_{\text{max}}^{qm} - \left(F_{L}\right)_{\text{max}}^{cl}}{F_{0}} \approx - \left[ 1 - \exp\left( - \frac{ \bar{\Lambda}^{-2}}{4} \right) \right] \approx - \frac{ \bar{\Lambda}^{-2}}{4} + \mathcal{O} \left( \bar{\Lambda}^{-2} \right).
\end{equation}
It appears that the difference between the quantum and classical forces is more prominent for smaller $\bar{\Lambda}$ showing $\bar{\Lambda}^{-2} \sim \hbar$ correlation. The quantum frictional process, however, becomes very similar to the classical case for $\bar{\Lambda}>1$, which marks the onset of more prominent $p=0$ and $p=1$ admixture. 

For $\eta \gtrsim 1$, the potential energy surface can have several local minima (Figures \ref{fig:2} (b,c)) causing stick-slip motion in both, the classical and quantum regimes (Figures \ref{fig:2}(g-i,j-$\ell$)). This type of motion was studied in detail in the adiabatic regime \cite{Zanca2018} and beyond \cite{le2025quantum} for the case of $\eta\sim 1$. The slipping time of the classical particle $t_{cl}^{slip} = T \left( {1}/{4} + {\eta}/{2 \pi} \right)$ is controlled by the corrugation parameter as found earlier \cite{Socoliuc2004, Gnecco2012, le2025quantum}. Regardless of $\bar{\Lambda}$, the classical lateral force is always a constant for a given $\eta$, and the results in Figure \ref{fig:3}(c, e) are consistent with the analytical expressions given in Equation \eqref{eqn:classical-lateral-approx}. 

The quantum motion for $\eta > 1$ is more complex due to its dependence upon $\bar{\Lambda}$. For example, for the case of $\eta=2.5$, Figure \ref{fig:2}(g-i) shows that there may not be stick-slip motion as evident for $\bar{\Lambda} = 0.2$, however, stick-slip motion occurs at $t/T\sim 0.5$ for $\bar{\Lambda} = 1$ and at $t/T\sim 0.65$ for $\bar{\Lambda} = 5$. We find that the depth of the potential minima play an important role for the LZ tunneling. In particular, shallow local minima specified by $\delta V / \hbar \Omega \leq \bar{\Lambda}^2 \left[ \eta + \sqrt{\eta^2 - u^2_{\eta}} - \frac{u^2_{\eta}}{2} \right] < 1$ do not host bound states (see Appendix \ref{appB} for details), thus they do not facilitate tunneling as also indicated by $P_{p=0}^{\text{min}} > 0.9$ for $\bar{\Lambda} < 1$ in Figure \ref{fig:3}(d). 


The lateral force correlates with the quantum trajectory of the nanoparticle. In the presence of LZ tunneling for, the maximal quantum force is  ${\left<F_L\right>_{\text{max}}^{qm}}/{F_{0}} \approx 0.5 - 1.0$ due to LZ tunneling, while the classical force is  ${\left<F_L\right>_{\text{max}}^{cl}}/{F_{0}} = 1$, as shown in Figure \ref{fig:3}(a,c,e). For all considered $\eta \gtrsim 1$, the quantum force approaches the classical limit for $\bar{\Lambda}>1$ as prominent LZ diabatic transitions cause enhanced disorder due to the admixture of neighboring energy levels in the system, as given in Figures \ref{fig:3}(d, f). This is consistent with the evolution of the maximal linear entropy of the nanoparticle quantum state, which increases  from $S_L^{\text{max}} \leq 0.01$ when $\bar{\Lambda} < 0.8$ to $S_L^{\text{max}} \geq 0.6$ when $\bar{\Lambda} \geq 1$, while the minimal ground state population drops down from $P_{p=0}^{\text{min}} \geq 0.9$ (no LZ tunneling) when $\bar{\Lambda} < 0.8$ to $P_{p=0}^{\text{min}} \approx 0$ (LZ tunneling present) when $\bar{\Lambda} \geq 1$.

Figures \ref{fig:2}(j-$\ell$) show that increasing $\eta$ to $\eta = 6.4$ leads to the quantum particle only oscillating around the center of the optical trap, as shown for $\bar{\Lambda} = 0.2$. In this case, the particle remains mostly stuck in the first potential minimum followed by a slip near the end of the period with slipping time $t^{slip}_{qm}\approx T$ for $\bar{\Lambda} = 1$. The particle also exhibits stick-slip motion when $\bar{\Lambda} = 5$ with the slipping time closer to the classical one $t^{slip}_{qm} \approx t^{slip}_{cl} \approx 1.3 T$. This is in agreement with Equation \eqref{eqn:quantum-lateral-3-threshold} which shows that for $\eta = 6.4$, LZ transitions cannot occur for $\bar{\Lambda} < 0.34$. This is also consistent with  $P^{\text{min}}_{p=0} < 0.5$ when  $\bar{\Lambda} < 0.4$ shown in Figure \ref{fig:3}(f). 

The lateral force in Figure \ref{fig:3}(e) has a similar behavior as for the case in Figure \ref{fig:3}(c). The quantum force is always smaller than the classical one with $\left<F_L\right>_{\text{max}}^{qm} \sim 0.01 $ for $\bar{\Lambda}<0.2$. The emergence of LZ tunneling enhances the force to $\left<F_L\right>_{\text{max}}^{qm} \sim 0.9 $ for $\bar{\Lambda}\sim 1$,and it approaches the classical value $\left<F_L\right>_{\text{max}}^{qm} \sim \left<F_L\right>_{\text{max}}^{cl} \sim 1.0 $ for larger  $\bar{\Lambda}$. The large $\eta$ and the emergence of LZ transitions also enhance the disorder in the system for $\bar{\Lambda}>0.5$, as found in the entropy results in Figure \ref{fig:3}(f).

The above results show that the corrugation parameter $\eta$ is closely related to the shape of the potential that controls the  stick-slip motion in both classical and quantum motions. The lateral force may experience a single stick-slip event as the particle passes the local minimum of the potential surface before transitioning to the next available minimum (typically for smaller $\eta$). The lateral force may also experience multi-slip events due to several minima of the potential surface before it becomes stuck again (typically for larger $\eta$). For example, as shown in Figures \ref{fig:S3}, \ref{fig:S4} (d-f) in Appendix \ref{appE}, the lateral force reaches its maximum $1.0 F_0$ not only at $t = 0.65 T$ but also at $t = 2.65 T$ when $\eta = 2.5$, indicating a single-slip feature. On the other hand, it reaches its maximum $1.0 F_0$ at $t \approx 1.3 T$ but its next local maximum is smaller, around $0.8 F_0$ at $t \approx 2.2 T$ when $\eta = 6.4$ (see Figures \ref{fig:S3}, \ref{fig:S4} (g-k) in Appendix \ref{appE}), indicating multi-slip features instead. The relative length ratio $\bar{\Lambda}$, on the other hand, is only relevant for the quantum motion since $\bar{\Lambda} \propto \hbar^{-1/2}$. Its role is to control the depth of the potential minima and the energy levels bound inside these minima, which in turn is important for the emergence  of LZ diabatic transitions and the associated regime of  quantum multi-slip motion. 

\subsection{\label{sec:eta}Modulations by changing the corrugation parameter $\eta$}

\begin{figure}[!ht]
    \centering
    \includegraphics[width=0.6\textwidth]{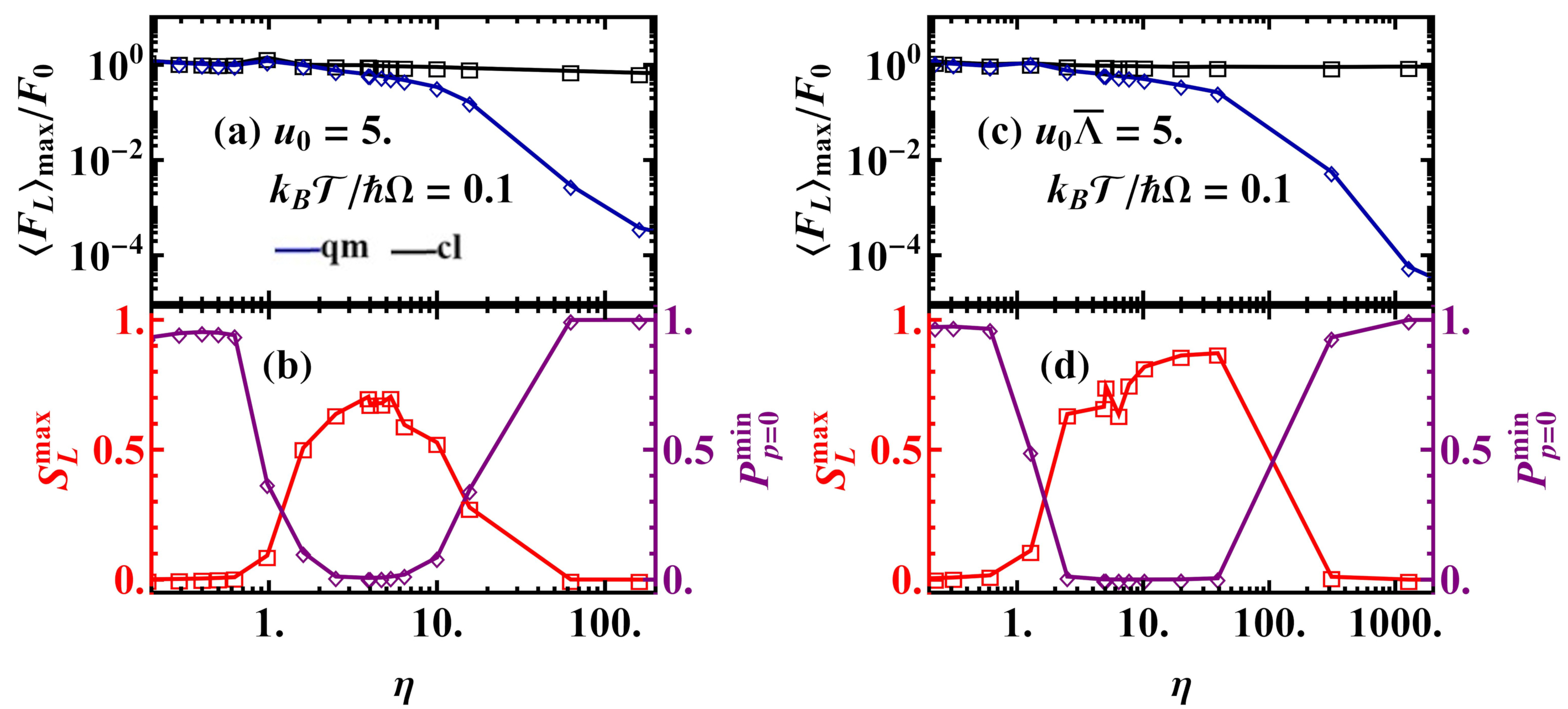}
    \caption{\label{fig:4} Classical (black) and quantum (blue) maximal lateral forces scaled by $F_0={\pi U_0}/{a}$ as a function of $\eta$ for: (a) fixed $u_0 = {U_0}/{\hbar \Omega}=5$, (c) fixed $u_0\bar{\Lambda}=5$. Maximal linear entropy $S_L^{\text{max}}$ (red) and minimal ground state 's population $P_{p=0}^{\text{min}}$  as a function of $\eta$ for: (b) fixed $u_0 = {U_0}/{\hbar \Omega}$ while tuning $\bar{\Lambda}$, (d) fixed $\Omega$ while tuning $a$.  The bath parameters are $\alpha = 2 \times 10^{-3}$, $\omega_c = 50$ $\Omega$ and $k_B \mathcal{T} = 0.1 $ $\hbar\Omega$. Also, $N_{\text{max}} = 200$ for all quantum and  $N_{\text{ran}}=200$ for all classical calculations.}
   
\end{figure}

Let us now investigate the frictional motion dependence as a function of the corrugation parameter. Taking into account that $\eta$ and $\bar{\Lambda}$ are composite parameters that depend on the physical characteristics $\Omega$, $a$, and $U_0$, we study two representative cases: tuning the lattice constant $a$ and frequency $\Omega$ by keeping $u_0 = {U_0}/{\hbar \Omega}$ fixed and tuning the lattice constant $a$ when keeping the frequency $\Omega$ fixed. Results from the numerical calculations are shown in Figure \ref{fig:4} for the classical and quantum frictional motion and its dependence upon $\eta$.


Under the first scenario with fixed $u_0$, we recognize that $U_0 \sim \bar{\Lambda}^{-2}$ and $\eta=u_0 \bar{\Lambda}^{-2}/2 \sim \bar{\Lambda}^{-2}$. This case is demonstrated in Figure \ref{fig:4}(a, b) for $u_0 = 5$. For $\eta<1$, the difference between the quantum and classical forces is minute, which  is consistent with the behavior for large $\bar{\Lambda}$ in  Equation \eqref{eqn:difference-1}. In this small $\eta$-- large $\bar{\Lambda}$ regime, the particle is mostly located in the ground state, which is also consistent with $P^\text{min }_{p=0}=(0.5, 0.9)$ for $\eta<1$ (Figure \ref{fig:4}(b)). In this range, the system also has little disorder from the thermal bath with $\text{max }S_L\sim 0$ as shown in Figure \ref{fig:4}(b). 

Increasing $\eta$ beyond $10$, however, leads to a reduced lateral force with an increasing disparity between the classical and quantum friction. As discussed previously in Equations \eqref{eqn:quantum-lateral-approx}-\eqref{eqn:classical-lateral-approx}, the classical force behaves as ${\left<F_L\right>_{\text{max}}^{cl}}/{F_{0}} \approx \sin \left( {2\pi}/{\eta} \right)$, while the quantum one is ${\left<F_L\right>_{\text{max}}^{qm}}/{F_{0}} \approx {2\pi}/{\eta \Omega T} + \exp{\left(-{\bar{\Lambda}^{-2}}/{4}\right)}$. The decrease in the classical force, especially for $\eta>10$, is attributed to the particle being stuck in the potential minimum until longer slipping times found to be beyond the first period $t_{cl}^{slip} = T \left( {1}/{4} + {\eta}/{2 \pi} \right) > T$. In the quantum motion, the particle is also stuck in the ground state exhibiting oscillations around the potential minimum. LZ tunneling transitions do not occur due to the shallow potential minimum for such large $\eta$.  The dominance of the ground state is also confirmed by Figure \ref{fig:4}(b), which shows that $P^\text{min }_{p=0} \approx 1$ and $S^\text{max }_L \approx 0$ for $\eta > 20$. 

In the intermediate range of $1<\eta<10$, both the classical and quantum particles exhibit stick-slip motion with ${\left<F_L\right>_{\text{max}}^{cl}}/{F_{0}} = 1$ and ${\left<F_L\right>_{\text{max}}^{qm}}/{F_{0}} \approx 0.6 - 0.8$ (Figure \ref{fig:4}(a)). In this regime, $P^\text{min }_{n=0} \approx 0$ and $S^\text{max }_L \approx 0.4$ indicating occurrence of LZ transitions and thermally induced disorder during the quantum motion (Figure \ref{fig:4}(b)).

Results for the quantum and classical motion for the second scenario under tuning the lattice constant $a$ while keeping $\Omega$ fixed are shown in Figure \ref{fig:4}(c, d). We note that in this case, $u_0 = {U_0}/{\hbar \Omega} \propto a^{-1} \propto \bar{\Lambda}^{-1}$ and hence the corrugation coefficient is scaled by $\eta = u_0 \bar{\Lambda}^{-2} /2 \propto \bar{\Lambda}^{-3}$. The corrugation parameter decreases as a function of the relative distance at a higher rate compared to the previously considered case of keeping $u_0$ fixed (Figure \ref{fig:4}(a,b)). The $\eta\sim \bar{\Lambda}^{-3}$ decay does not introduce qualitative changes in the quantum and classical frictional motion. Figure \ref{fig:4}(c,d) shows that there are quantitative differences when compared to the slower $\eta\sim \bar{\Lambda}^{-2}$ dependence corresponding to  Figure \ref{fig:4}(a,b). The mechanisms for the frictional force and released heat behavior in the different $\eta$ regions are the same in both cases. The no stick-slip regimes are found for $\eta<1$ and $\eta>40$, while $1\lesssim\eta<10$ corresponds to stick-slip motion although the LZ tunneling renders smaller quantum frictional force compared to the classical one. 

\begin{figure}[!ht]
    \centering
    \includegraphics[width=0.9\textwidth]{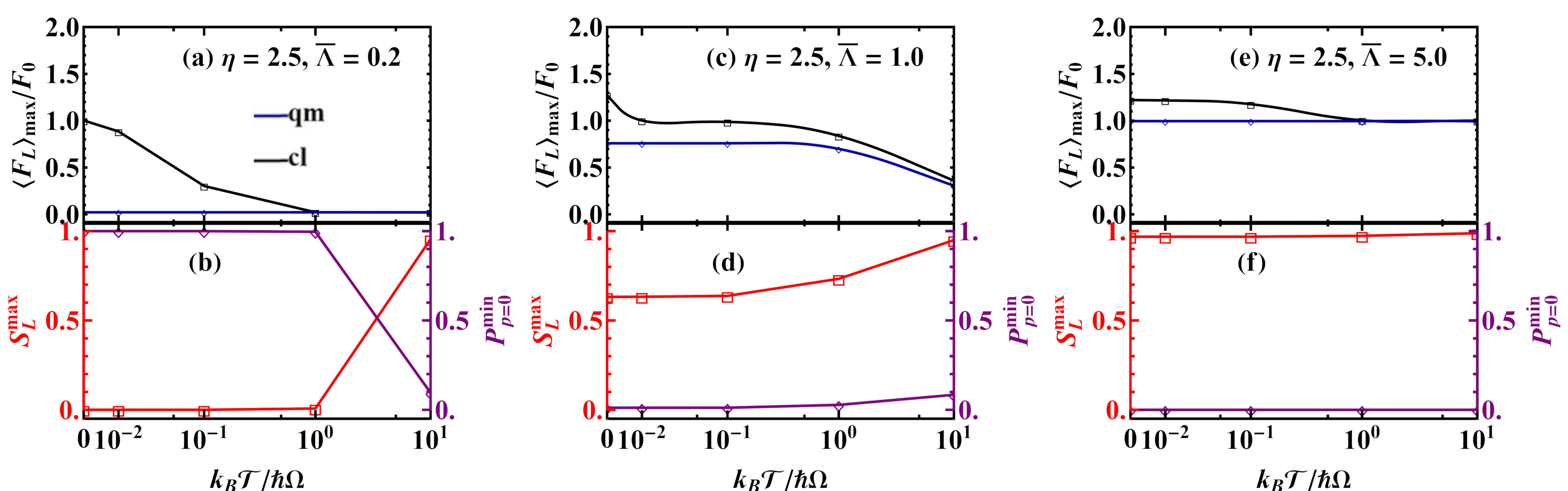}
    \caption{\label{fig:5} Classical (black) and quantum (blue) maximal lateral forces scaled by $F_0={\pi U_0}/{a}$ as a function of $k_B \mathcal{T}/\hbar \Omega$ for: (a) $\eta=2.5, \bar{\Lambda} = 0.2$; (b) $\eta=2.5, \bar{\Lambda} = 1.0$; (c) $\eta=2.5, \bar{\Lambda} = 5.0$. Maximal linear entropy $S_L^{\text{max}}$ (red) and minimal ground state  population $P_{p=0}^{\text{min}}$ (purple) as a function of $k_B \mathcal{T}/\hbar \Omega$ for: (d) $\eta=2.5, \bar{\Lambda} = 0.2$; (e) $\eta=2.5, \bar{\Lambda} = 1.0$; (f) $\eta=2.5, \bar{\Lambda} = 5.0$. The cutoff frequency $\omega_c = 50 \Omega$ and $N_{\text{max}} = 200$ for all quantum calculations. For the classical calculations, $N_{\text{ran}} = 4 \times 10^6$ for $\mathcal{T} = 10 \hbar \Omega$ in (a), while $N_{\text{ran}} = 200$ for the rest of the cases.}
\end{figure}

We now examine how temperature affects the nanoscaled friction. In Figure \ref{fig:5}(a-c), we show results for the force as a function of  $k_B \mathcal{T}/\hbar \Omega$ for different $\eta$ and $\bar{\Lambda}$ values. The linear entropy and population of the ground state are also shown (Figure \ref{fig:5}(d-f)).  
The quantum force for $\bar{\Lambda} = 0.2$ remains practically unchanged over the entire temperature region (Figure \ref{fig:5}(a)). This is associated with lack of LZ diabatic tunneling as discussed in Section \ref{sec:Lambda}. For $\bar{\Lambda} = 1$, however, LZ tunneling is possible which is further stimulated by higher temperatures. This causes the particle to slip at earlier time resulting in  reduction of the quantum frictional force (Figure \ref{fig:5}(c,d)). The quantum force also remains unchanged for $\bar{\Lambda}=5$ for all temperatures (Figure \ref{fig:5}(e)). In this case, the potential minimum is very deep ($\delta V/\hbar \Omega \approx 39$), thus many thermally populated states are found inside it. As a result, many LZ transitions occur, which is also consistent with  $P_{p=0}^{\text{max}} \approx 0$ as shown in Figure \ref{fig:5}(f). Consequently, the quantum force approaches the classical limit $\left< F_L \right>_{\text{max}}^{qm} \approx 1.0 F_0$ as shown in Figure \ref{fig:5}(e).

In the classical regime, the environmental temperature affects the system differently. According to Equation \eqref{eqn:classical}, $\mathcal{T}$ appears solely in the random force $\xi_{ran}$, whose magnitude is approximately $\sqrt{ \left({\alpha (\Omega T)^3}/{2\pi \bar{\Lambda}^4}\right) \left({k_B \mathcal{T}}/{\hbar \Omega}\right) } \propto \bar{\Lambda}^{-2} \sqrt{{k_B \mathcal{T}}/{\hbar \Omega} }$, and mostly affects on the deviation of the force rather than the average value of the lateral force. Figure \ref{fig:5} shows that 
$\left< F_L \right>_{\text{max}}^{qm} \rightarrow \left< F_L \right>_{\text{max}}^{cl}$ as $k_B \mathcal{T}$ increases ensuring the validity of the high temperature correspondence principle of our model.

\section{\label{sec:conc}Outlook, Conclusions}

Nanoscaled friction is complex with diverse behavior of the frictional force. The present study systematically analyzes quantum mechanical and classical effects within the Prandtl-Tomlinson model, which serves as a central framework in tribology research. By focusing on the changes of lateral force  when modulating different model parameters, we reveal that the corrugation coefficient $\eta$, which is associated with the shape of the effective potential, controls the occurrence of stick-slip motion in the classical regime of motion. For the quantum friction, however, beside  $\eta$ the relative length ratio $\bar{\Lambda}$ between the lattice spacing $a$ of periodic potential and the characteristic length $\ell$ of the harmonic potential is also important. We find that the quantum stick-slip can be controlled by both parameters as  LZ tunneling can be affected by   $\eta$ and $\bar{\Lambda}$. 

Our study shows for the first time how system properties, such as interaction strength, lattice spacing, and  frequency of the driving optical trap, can be tuned through the composite corrugation coefficient $\eta$ and relative length ratio $\bar{\Lambda}$ to achieve different regimes of motion beyond stick-slip. Depending on $\eta$ and $\bar{\Lambda}$, it is possible that the quantum particle can experience not only a single stick-slip event, but multi-slip motion or motion with no LZ tunneling. As a result, the quantum and classical forces can differ significantly.  
This is especially important for understanding how to enhance or inhibit the frictional force by manipulating various physical characteristics of the system. Uncovering that the quantum motion can be controlled by two composite parameters as opposed to the classical motion shows that LZ tunneling brings complexity to nanoscaled friction with additional opportunities for control. 

 Our findings offer a theoretical foundation for future experimental investigations into quantum effects in stick-slip motion. Our study may also be suitable for the development of future neural network models capable of learning both classical \cite{Shabani2025} and quantum frictional behavior at the nanoscale. Experimental platforms such as cold atoms and ions in optical lattices, along with time-resolved LZ tunneling in periodic potentials, already demonstrated in laboratory settings \cite{Gangloff2015, Counts2017, Bonvin2024, Zenesini2009}, may serve as promising systems to observe frictional reduction in the quantum regime. 

\begin{acknowledgments}
We acknowledge financial support from the US National Science Foundation under grant No. 2306203.
\end{acknowledgments}

\appendix

\renewcommand{\thefigure}{\Alph{section}\arabic{figure}}

\section{\label{appA} Analytical expressions for the classical motion }
\setcounter{figure}{0}
Here we provide analytical derivations of the lateral force when the dissipation in the system is neglected. This corresponds to a regime of low viscosity for which  $\bar{\Lambda} \gg \sqrt{2\pi \alpha \Omega T}$ and the last two terms of the stochastic Newton 's equation \eqref{eqn:classical} can be neglected.

\paragraph{The case of $\eta < 1$}: In this case, the potential energy surface is a single-well potential whose minimum is located at the center of the driving trap (see Figure \ref{fig:2}(a)). The particle's motion is oscillatory around it with $\frac{x}{a} \approx \frac{t}{T} + \frac{\delta x}{a}$ where $|\delta x| \ll a$ is the displacement around the potential minimum. Inserting this ansatz  into  Equation \eqref{eqn:classical} and expanding up to first order in $\delta x$, we obtain  Newton's equation of motion for $\delta x$ from which we can approximately solve the trajectory of the particle as
\begin{eqnarray}
    \label{eqn:classical-eta-1}
    \dfrac{x}{a} \approx \dfrac{t}{T} -\dfrac{1}{\Omega T} \sin \left( \Omega t \right) + \dfrac{\eta}{2 \pi} \sin \left( \frac{2\pi t}{T} \right)
\end{eqnarray}
The lateral force in Equation \eqref{eqn:classial-lateral} is then obtained as 
\begin{eqnarray}
    \label{eqn:classial-lateral-1}
    \frac{\left<F_L\right>^{cl}(t)}{F_{0}} = \frac{2\pi}{\eta} \left( \frac{t}{T} - \frac{x}{a}\right) \approx \left[ \frac{2\pi}{\eta \Omega T} \sin (\Omega t) + \sin \left( - \frac{2\pi t}{T} \right) \right] \leq 1+ \dfrac{2\pi}{\eta \Omega T}  \Rightarrow \frac{\left<F_L\right>_{\text{max}}^{cl}}{F_{0}} \approx   1+ \dfrac{2\pi}{\eta \Omega T}. 
\end{eqnarray}

\paragraph{The case of  $1 \lesssim  \eta$}: 
In this case, the potential energy surface forms a multi-well shape (see Figure \ref{fig:2} (b)). During its motion, however, the particle remains in the first local minimum of its quasi-static potential surface where it feels almost zero net force i.e. the first term on the right-hand side of Equation \eqref{eqn:classical} is approximately zero
\begin{eqnarray}
    \label{eqn:classical-2}
    \dfrac{x}{a} - \dfrac{t}{T} + \dfrac{\eta}{2\pi} \sin \left( \dfrac{2\pi x}{a} \right) \approx 0, \quad \text{when } 0 \leq \dfrac{t}{T} \leq \frac{1}{4} + \frac{\eta}{2\pi},
\end{eqnarray}
At the slipping time $t_{cl}^{slip} = T \left( {1}/{4} + {\eta}/{2\pi} \right) < T$, we have $\frac{x}{a} = \frac{1}{4}$ and the lateral force in Equation \eqref{eqn:classial-lateral} reaches its maximum \cite{Socoliuc2004, Gnecco2012, le2025quantum}
\begin{eqnarray}
    \label{eqn:classial-lateral-2}
    \frac{\left<F_L\right>_{\text{max}}^{cl}}{F_{0}}  = \frac{2\pi}{\eta} \left( \frac{t}{T} - \frac{x}{a}\right)_{\text{max}} \approx \left[ \sin \left( \frac{2\pi x}{a}\right) \right]_{\text{max}} \approx 1.
\end{eqnarray}

\section{\label{appB} Analytical expressions for the quantum motion }
\setcounter{figure}{0}
Neglecting dissipation in the quantum mechanical motion also yields useful expressions.
\paragraph{For $\eta < 1$}: In this case, the total potential surface does not form local minima and LZ diabatic transitions between states do not occur, i.e. the quantum motion does not experience stick-slip. For this situation, a two-level approximation is sufficient to qualitatively analyze the time evolution of the system. Focusing only on the lowest two levels ($p = 0$ and $p = 1$) and neglecting the effect of the thermal bath, we can write the density matrix operator of the system in the time-dependent moving harmonic oscillator basis set $ \left\{ \left| 0^{(0)} (t)  \right>, \left| 1^{(0)} (t)  \right>  \right\} $ as
\begin{eqnarray}
    \label{eqn:quantum-density-1}
    \hat{\rho}_S (t) && = \left| c_0 (t) \right|^2 \left| 0^{(0)} (t)  \right> \left< 0^{(0)} (t) \right| + \left| c_1 (t) \right|^2 \left| 1^{(0)} (t)  \right> \left< 1^{(0)} (t)  \right| \nonumber\\
    && + c_0(t) c_1^* (t) \left| 0^{(0)} (t)  \right> \left< 1^{(0)} (t)  \right| + c_0^*(t) c_1 (t) \left| 1^{(0)} (t)  \right> \left< 0^{(0)} (t)  \right| .
\end{eqnarray}
Substituting the above expression into 
 Equations \eqref{eqn:quantum-1}-\eqref{eqn:quantum-3} we obtain the $c_{0,1} (t)$ coefficients,
\begin{eqnarray}
    c_0 (t) && \approx 2 \cos \left( \frac{\varphi(t)}{2} \right) - \left( \frac{\eta \Omega T}{2\pi} \right) e^{-\frac{\bar{\Lambda}^{-2}}{4}} e^{-i \left[ \frac{\varphi(t)}{2} + \Omega \sqrt{1 + 2 \bar{\Lambda}^2 \left( \frac{2\pi}{\Omega T} \right)^2 } t \right] } \sin \left( \frac{2\pi t}{T} \right), \\
    c_1 (t) && \approx \frac{\bar{\Lambda}}{\sqrt{2}}  \left( \frac{2\pi}{\Omega T} \right) \left[ 2 \sin \left( \frac{\varphi(t)}{2} \right) - \left( \frac{\eta \Omega T}{2\pi} \right) e^{-\frac{\bar{\Lambda}^{-2}}{4}} e^{-i \left[ \frac{\varphi(t)}{2} + \Omega \sqrt{1 + 2 \bar{\Lambda}^2 \left( \frac{2\pi}{\Omega T} \right)^2 } t + \frac{\pi}{2} \right] } \sin \left( \frac{2\pi t}{T} \right) \right],
\end{eqnarray}
with $\varphi (t) \approx \Omega \sqrt{1 + 2 \bar{\Lambda}^2 \left( \frac{2\pi}{\Omega T} \right)^2 } t + \frac{\left( \frac{\eta \Omega T}{2\pi} \right) e^{-\frac{\bar{\Lambda}^{-2}}{4}}}{2\sqrt{1 + 2 \bar{\Lambda}^2 \left( \frac{2\pi}{\Omega T} \right)^2 } } \sin \left( \frac{2\pi t}{T} \right)$.

Plugging the so-obtained density matrix $\hat{\rho}_S$ (Equation \eqref{eqn:quantum-density-1}) into Equation \eqref{eqn:quantum-lateral}, we can determine the lateral force as
\begin{equation}
    \label{eqn:quantum-lateral-1}
    \frac{\left< F_L \right>^{qm} (t)}{F_{0}} = -\frac{\sqrt{2}}{\eta} \frac{2\pi \ell}{a} \Re\left[ c_0 (t) c^*_1(t) \right] \approx \dfrac{2\pi}{\eta \Omega T} \sin \left( - \varphi(t) \right) + e^{-\frac{\bar{\Lambda}^{-2}}{4}} \sin\left[ \varphi(t) + \Omega \sqrt{1 + 2 \bar{\Lambda}^2 \left( \frac{2\pi}{\Omega T} \right)^2 } t \right] \sin \left( \frac{2\pi t}{T} \right).
\end{equation}
Taking that the $\sin$-function is less than 1, we find that the lateral force is
\begin{eqnarray}
    \label{eqn:quantum-lateral-2}
   \frac{\left< F_L \right>^{qm} (t)}{F_{0}} \leq  \dfrac{2\pi}{\eta \Omega T} + e^{-\frac{\bar{\Lambda}^{-2}}{4}} \quad  \Rightarrow \frac{\left< F_L \right>^{qm}_{\text{max}}}{F_{0}} \approx   \dfrac{2\pi}{\eta \Omega T} + e^{-\frac{\bar{\Lambda}^{-2}}{4}}. 
\end{eqnarray}

\paragraph{For $\eta \gtrsim 1$:} In this case, the effective potential forms a multi-well shape. However, unlike the classical motion, the quantum motion also depends on the relative length ratio $\bar{\Lambda}$, since this parameter controls the dependence of anti-crossing levels on the depth of the local minima of the potential \cite{Le2018}.  

We find that in the middle of the period $t = \frac{T}{2}$, the effective potential near the center of driving trap $\frac{x}{a} \approx \frac{t}{T}$ forms at least two minima whose depth is $\delta V = V_{\text{max}} - V_{\text{min}} = \hbar \Omega \bar{\Lambda}^2 \left[ \eta + \sqrt{\eta^2 - u^2_{\eta}} - \frac{u^2_{\eta}}{2} \right]$ where $u_{\eta}$ is the solution of equation $u = \eta \sin u$ within the interval $\left( 0 , \pi \right)$. If these minima are shallow with $\delta V \ll \hbar \Omega$, there are no energy levels inside the minima. LZ diabatic transitions do not occur while the particle is in its ground state. The maximal lateral force is then similar to the  case of $\eta < 1$ 
\begin{eqnarray}
    \label{eqn:quantum-lateral-3}
    \frac{\left< F_L \right>^{qm}_{\text{max}}}{F_0} \approx   \dfrac{2\pi}{\eta \Omega T} + e^{-\frac{\bar{\Lambda}^{-2}}{4}},
\end{eqnarray}
for
\begin{eqnarray}
    \label{eqn:quantum-lateral-3-threshold}
    \bar{\Lambda}^2 \ll \frac{1}{\eta + \sqrt{\eta^2 - u^2_{\eta}} - \frac{u^2_{\eta}}{2}}
\end{eqnarray}
For very large $\eta \gg 1$, $u_{\eta} \approx \frac{\eta \pi}{\eta+1}$, and then the threshold \eqref{eqn:quantum-lateral-3-threshold} reads
\begin{eqnarray}
    \label{eqn:quantum-lateral-3-threshold-large}
\bar{\Lambda}^2 \ll \frac{1}{\eta \left(2 - \frac{\pi^2}{2(1+\eta)} \right)}
\end{eqnarray}

We can also obtain an approximate analytical expression when $\bar{\Lambda}\sim 1$, in which case LZ tunneling is possible since some energy levels are located inside the potential well (see Figures. \ref{fig:2}(b,c) for example). In this case, the probability transition between $p^{\text{th}}$ and $(p+1)^{\text{th}}$ is velocity dependent $P_{p,p+1} = \exp\left( - \frac{\pi |E_{p+1} - E_p|^2}{2 |\alpha_{p+1} - \alpha_p| v} \right)$ where $\alpha_p = \frac{\partial E_{p}}{\partial x_c}$ is the slope of eigenenergy $E_p$ with respect to the center of driving trap $x_c = vt$ \cite{Landau1932, Zener1932, Zanca2018, le2025quantum}. As long as the potential energy surface does not evolve to a different shape,  the slipping time $t_{p}^{slip}$ for the $p^{th}$ eigenstate  is always smaller than the classical one i.e. $t_{p}^{slip} < t_{cl}^{slip} =  T \left(\frac{1}{4} + \frac{\eta}{2 \pi} \right)$ \cite{le2025quantum}. As the motion evolves, more LZ transitions occur, which means that the particle is stuck in their minima for longer times. The particle starts slipping around
\begin{eqnarray}
    \label{eqn:quantum-lateral-4}
    t^{slip}_{qm} \approx t_0^{slip} + \sum_{\substack{n>0\\t_{p+1}^{slip} \leq T}} P_{0,1} P_{1,2} P_{2,3} \ldots P_{p,p+1} \left( t_{p+1}^{slip} - t_{p}^{slip} \right).
\end{eqnarray}
Since all $P_{p,p+1}<1$, we find
\begin{eqnarray}
    \label{eqn:quantum-lateral-4b}
    t^{slip}_{qm} < t_0^{slip} + \sum_{\substack{n>0\\t_{p+1}^{slip} \leq T}} \left( t_{p+1}^{slip} - t_{p}^{slip} \right) = \min \left[ t_{p+1}^{slip} , T \right] <  \min \left[ t_{cl}^{slip}, T \right].
\end{eqnarray}
The maximal lateral force in Equation \eqref{eqn:quantum-lateral} is essentially  proportional to the slipping time $t^{slip}_{qm}$
\begin{eqnarray}
    \label{eqn:quantum-lateral-5}
    \frac{\left< F_L \right>^{qm}_{\text{max}}}{F_0} \approx  \frac{t_{qm}^{slip}}{t_{cl}^{slip}}  < \frac{\min \left[ t_{cl}^{slip}, T \right]}{t_{cl}^{slip}},
\end{eqnarray}
which is always smaller than the classical one, 
\begin{eqnarray}
    \label{eqn:quantum-lateral-6}
    & \frac{\left< F_L \right>^{qm}_{\text{max}}}{F_0} \sim \text{const} < 1 & \text{ when } 1 \leq \eta .
\end{eqnarray}

\section{\label{appC}Convergence test for the quantum calculations and standard deviation for the lateral force}
\setcounter{figure}{0}
Here we show our results for computations of the quantum force for different basis sets specified by the size of the matrix $N_{\text{max}}$. We find that for $\bar{\Lambda} = 1$ and $\eta = 2.5, 6.4$ increasing the cutoff beyond the original $N_{\text{max}}=25$ does not lead to changes in $\left<F_L \right>_{\text{max}}$. Convergence within six significant digits is achieved with $N_{\text{max}}=300$ for $\eta=2.5, \bar{\Lambda}=8$ and $N_{\text{max}}=500$ for $\eta=2.5, \bar{\Lambda}=10$. For the case of for $\eta=6.4, \bar{\Lambda}=5$ convergence of  5\% is obtained with $N_{\text{max}} = 650$. For  $\eta=0.1$, the quantum force does not change for all considered $\Lambda$ upon increasing $N_{\text{max}}$ suggesting that the initial $N_{\text{max}}=25$ is sufficient.


\textcolor{red}{
\begin{figure}[H]
    \centering
    \includegraphics[width=0.5\textwidth]{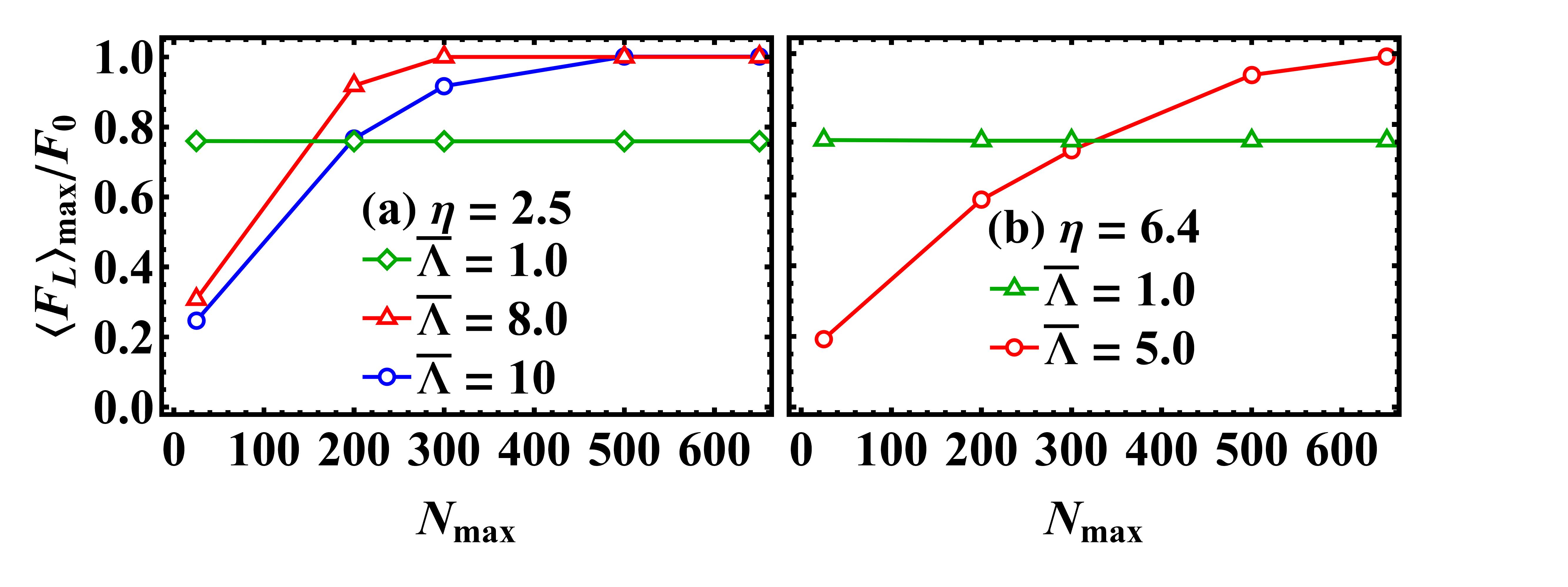}
    \caption{\label{fig:S1} Quantum maximal lateral force normalized by $F_0$ versus size of basis set  $N_{\text{max}}$ for (a) $\eta = 2.5$ with $\bar{\Lambda} = 1.0, 8.0$ and $10$ and (b) $\eta = 6.4$ with $\bar{\Lambda} = 1.0$ and $5.0$.}
\end{figure}
}

\begin{figure}[!ht]
    \centering
    \includegraphics[width=0.75\textwidth]{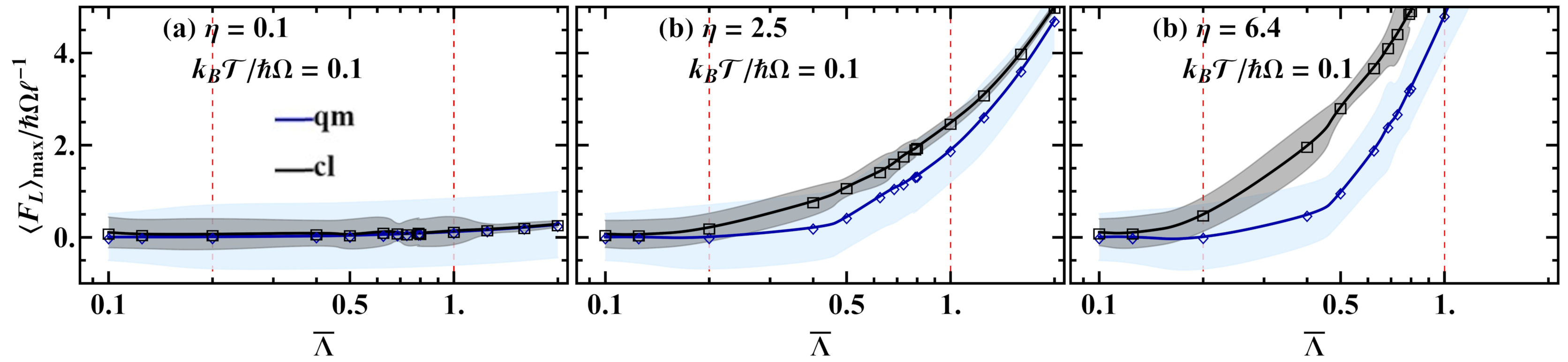}
    \caption{\label{fig:S2} Classical (black) and quantum (blue) maximal lateral force $\left<F_L\right>_{\text{max}}/\hbar\Omega\ell^{-1}$ versus $\bar{\Lambda}$ for (a) $\eta = 0.1$, (b) $\eta = 2.5$ and (c) $\eta = 6.4$, with shaded bands representing their standard deviations $\sigma \left<F_L\right>_{\text{max}}/\hbar\Omega\ell^{-1}$.}
\end{figure}

We also evaluate the standard deviation of the maximal quantum lateral force, $\sigma \left<F_L\right>_{\text{max}}=\left. \sqrt{\left<F_L^2\right> - \left<F_L\right>^2}\right|_{\left<F_L\right> = \left<F_L\right>_{\text{max}}}$ for the lateral force operator $\hat{F}_L$ from the known density matrix $\hat{\rho}$ when the lateral force reaches its maximum in in the $t/T=[0,3]$ range. The data is shown for $N_{\text{max}}=200$. Similarly, $\sigma \left<F_L\right>_{\text{max}}$ is evaluated for the classical force over $N_{\text{ran}} = 200$ independent classical simulations. Figure \ref{fig:S2} illustrates the maximal lateral force $\left<F_L\right>_{\text{max}}/\hbar\Omega\ell^{-1}$ versus $\bar{\Lambda}$ with shaded bands representing their standard deviations $\sigma \left<F_L\right>_{\text{max}}/\hbar\Omega\ell^{-1}$.

\section{\label{appE}Time evolution for the lateral force}
\setcounter{figure}{0}

Figures \ref{fig:S3} and \ref{fig:S4} represent the time dependence of the quantum and classical lateral forces $\left<F_L\right>^{\text{qm,cl}}(t)/F_0$ for $k_B \mathcal{T} = 0.01$ $\hbar\Omega$ and $k_B \mathcal{T} = 0.1$ $\hbar\Omega$, respectively. The results are shown for $(\eta, \bar{\Lambda})$ as those in Figure \ref{fig:2}(d-l).

\textcolor{red}{
\begin{figure}[!ht]
    \centering
    \includegraphics[width=0.75\textwidth]{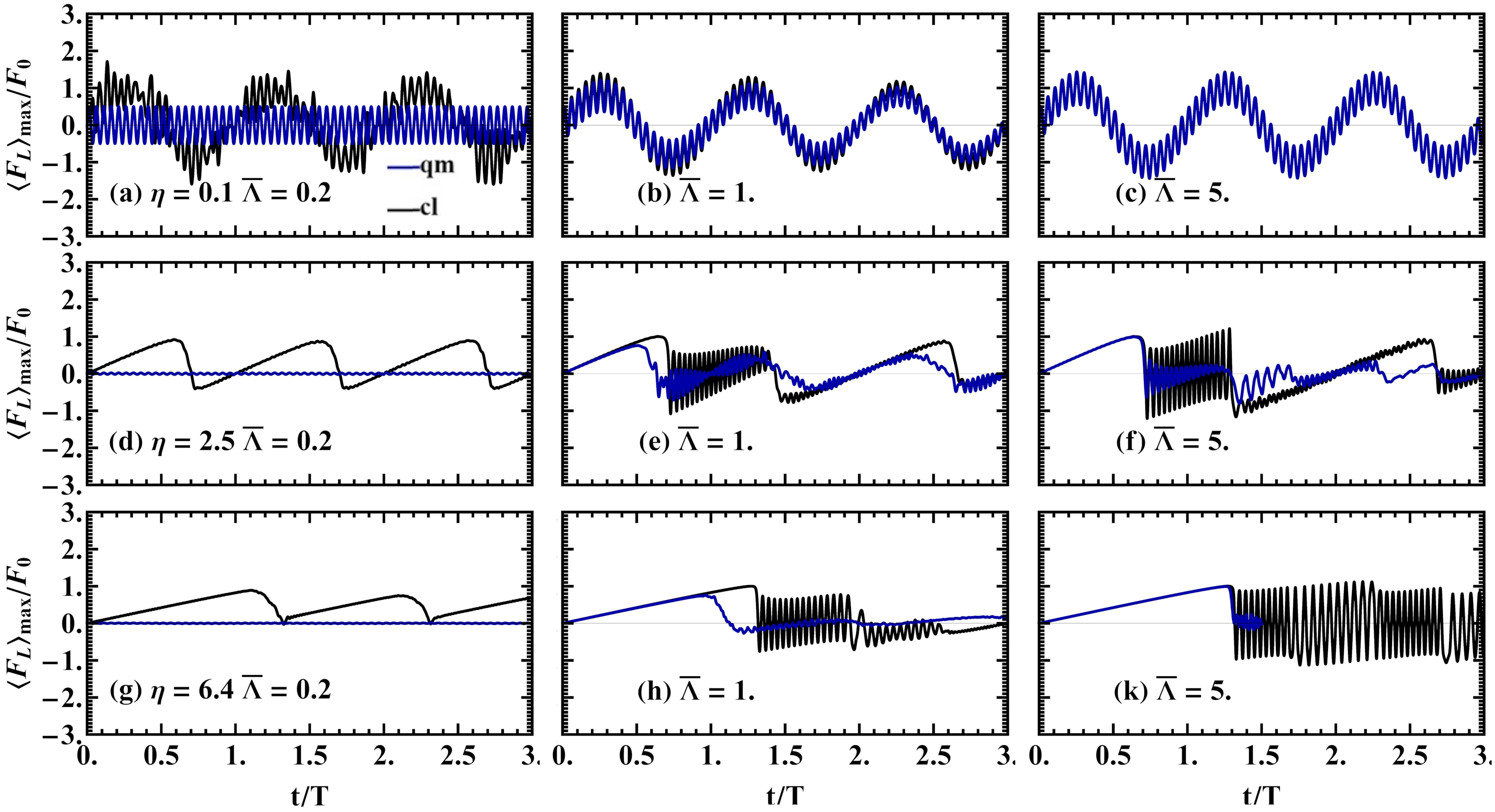}
    \caption{\label{fig:S3} Classical (black) and quantum (blue) lateral force normalized by $F_0$ i.e. $\left<F_L\right>/F_0$ over time $t/T$ for $\eta=0.1$ at (a) $\bar{\Lambda} = 0.2$, (b) $\bar{\Lambda} = 1$, (c) $\bar{\Lambda} = 5$; $\eta=2.5$ at (d) $\bar{\Lambda} = 0.2$, (e) $\bar{\Lambda} = 1$, (f) $\bar{\Lambda} = 5$; and $\eta=6.4$ at (g) $\bar{\Lambda} = 0.2$, (h) $\bar{\Lambda} = 1$, (i) $\bar{\Lambda} = 5$. The bath parameters are $\alpha = 2 \times 10^{-3}$, $\omega_c = 50\Omega$ and $k_B \mathcal{T} = 0.01 \hbar\Omega$. Note that in (k), quantum simulation can be performed only up to $t/T = 1.5$.}
\end{figure}
}
\textcolor{red}{
\begin{figure}[!ht]
    \centering
    \includegraphics[width=0.75\textwidth]{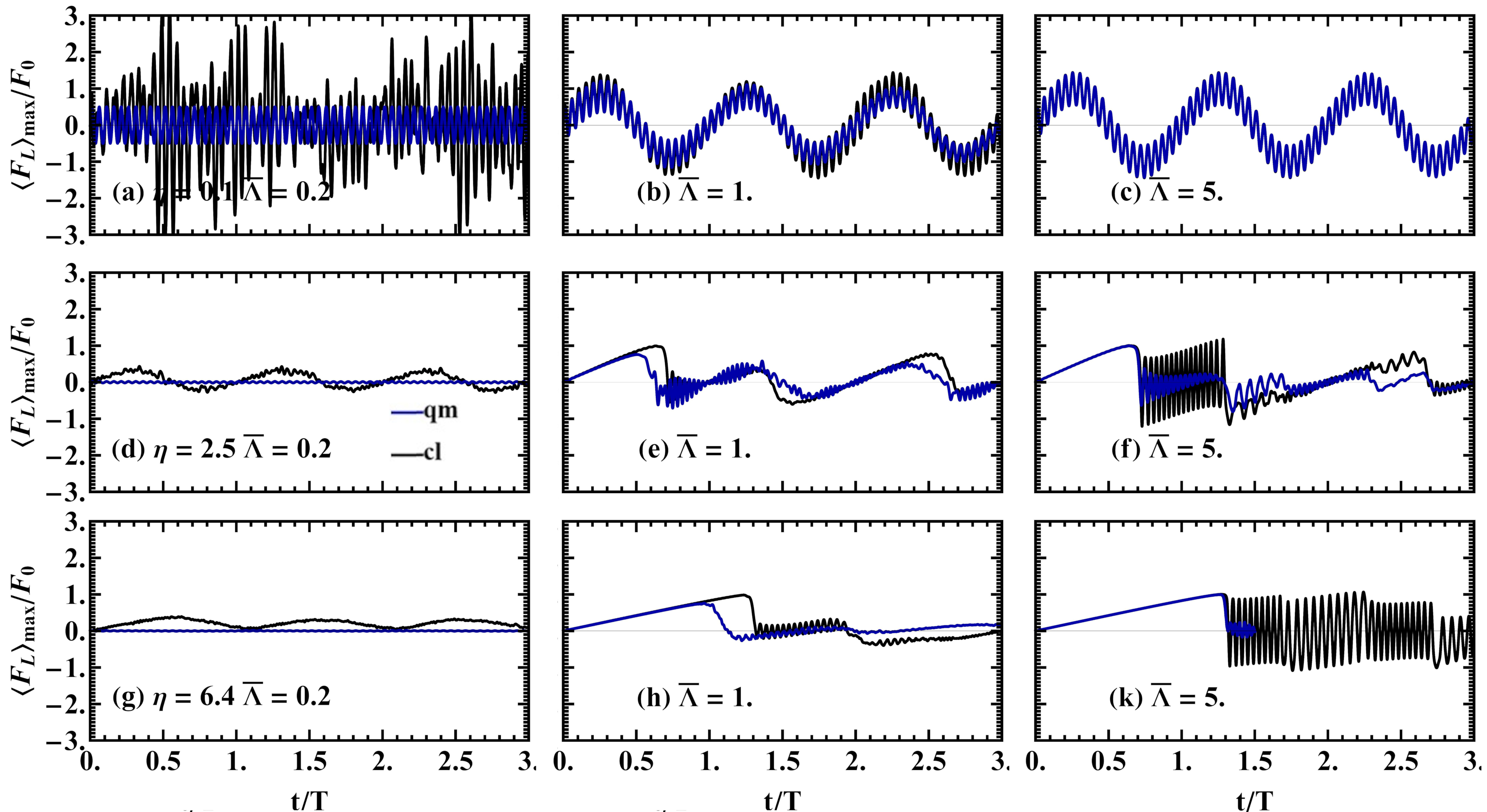}
    \caption{\label{fig:S4} Classical (black) and quantum (blue) lateral force normalized by $F_0$ i.e. $\left<F_L\right>/F_0$ over time $t/T$ for $\eta=0.1$ at (a) $\bar{\Lambda} = 0.2$, (b) $\bar{\Lambda} = 1$, (c) $\bar{\Lambda} = 5$; $\eta=2.5$ at (d) $\bar{\Lambda} = 0.2$, (e) $\bar{\Lambda} = 1$, (f) $\bar{\Lambda} = 5$; and $\eta=6.4$ at (g) $\bar{\Lambda} = 0.2$, (h) $\bar{\Lambda} = 1$, (i) $\bar{\Lambda} = 5$. The bath parameters are $\alpha = 2 \times 10^{-3}$, $\omega_c = 50\Omega$ and $k_B \mathcal{T} = 0.1 \hbar\Omega$.}
\end{figure}
}

\newpage
\def\bibsection{\section*{References}}

\bibliography{refs}

\end{document}